\def\beg{\begin{equation}}
\def\eeq{\end{equation}}
\documentstyle[12pt]{article}
\begin{document}
\begin{center}
{\Large{\bf Fractional Charge Experiments: What quantity is measured 
in the quantum Hall effect or calculated by Laughlin?}}
\vskip0.35cm
{\bf Keshav N. Shrivastava}
\vskip0.25cm
{\it School of Physics, University of Hyderabad,\\
Hyderabad  500046, India}
\end{center}

We have examined the experiments performed by Goldman and Su, 
de-Picciotto et al, Saminadayar et al and Comforti et al, in 
which it is claimed that a fractional charge of e/3 is found. 
In all of the measurements, the quantity measured is the product 
of the charge and 
the magnetic field but not the charge. It is possible to interpret 
that charge per unit area has been measured where area is the
 square of the magnetic length. This type of correction to
 Laughlin's result does not affect the exactness of the calculation.
 Anderson has suggested the extension of Laughlin's state to 
particles of charge 2/m or 3/m with m=odd integer. We find that 
the quasiparticle charge is given by angular momenta, 
$e_{eff}/e$ =${{\it l} +(1/2)\pm s\over 2{\it l}+1}$ which agrees 
with the data. Therefore, Laughlin's 1/odd becomes an angular 
momemtum so that the charge depends on spin, s, and zero charge
 has spin 1/2.

\vskip1.0cm
Corresponding author: keshav@mailaps.org\\
Fax: +91-40-2301 0145.Phone: 2301 0811.
\vskip1.0cm

\noindent {\bf 1.~ Introduction}

     Several authors have claimed that they measure a fractional 
charge of e/3. The question here is,  that whether such a charge 
occurs in vacuum in absence of a magnetic field? If not, then what 
is the 
quantity which is measured and yet, it is thought to be charge. If 
the electrons by themselves, due to Coulomb repulsion, produce a 
fractional charge then why in the first instance, it comes out to 
be e/3 and not e/2 or e/10 and how the charge is conserved. Then 
what determines that it should split into three equal parts and 
what happened to the mass? 
Is it possible to have three parts with one of mass $m$ and the 
other two of mass zero each or all parts have mass $m/3$ each. 
Is it possible to split the charge but not the mass? All such 
questions can best be answered by reexamination of the experimental 
data. We have, therefore, reexamined all of the data and find that
 in all of the experiments, the quantity measured is not the charge
 but the product of the charge and the magnetic field. This 
necessitates, rechecking of the Laughlin's theory of fractional
 charge. In this case also, it is found that charge per unit area
 is the quantity where the area is $\phi_o/B$. Hence, the
 experimentalists, have not measured the pure charge.

     In this paper, we explain all of the experimental data in
 relation to the quantity measured. It is found that the product
 of the charge
 and the magnetic field has been measured.

\noindent{\bf 2.~~Experiments}

     {\bf (a)}: Goldman and Su[1-3] claim to measure the fractional
 charge.
We have no objection to the measurement of fractional charge but it 
will be nice to know the quantity which is being measured. The 
resonant tunneling (RT) conductance peak is measured as a function 
of (a) magnetic field and (b) the back gate voltage and it is claimed
 that
 `` a combination of these two measurements for a given $\nu$ in 
the quantum Hall effect constitutes a direct measurement of the 
charge of the tunneling quasiparticles". At some condition between 
$B^m$ and $V_{BG}^m$, the lowest unoccupied state is aligned with
 magnetic moment $\mu$. At these values of field and voltage, the
 resonant tunneling is possible with a peak in the tunneling 
conductance, $G_{tun}$. At a fixed voltage $V_{BG}$, when $B$ is 
increased, the area $S_m$ decreases because the product $B.S_m$ is
 a constant due to quantization condition,
\beg
B^m.S_m=m\phi_o.
\eeq
When $m$ changes to $m+1$, the field should also change,
\beg
B^{m+1}.S_{m+1}=(m+1)\phi_o.
\eeq
The difference between the two fields should be quantized,
\beg
B^{m+1} - B^m= \Delta B.
\eeq
Apparently, there is no objection to the quantization of the
 difference but the difference can not have the same area as $S_m$.
  In any case, if we substract the above two equations, we find,
\beg
B^{m+1}S_{m+1}-B^mS_m=\phi_o.
\eeq
Goldman and Su write this equation as
\beg
\Delta B S_m=\phi_o.
\eeq
This is possible if,
\beg
S_{m+1}\simeq S_m
\eeq
so that
\beg
(B^{m+1}-B^m)S_m=\phi_o
\eeq
but it is dangerous to take such an approximation. Let us proceed 
with what we have. The voltage is written in terms of charge and 
capacitance as,
\beg
\Delta V_{BG}={q\over CS_m}
\eeq
where the capacitance is,
\beg
C={\epsilon \epsilon_o\over d_{BG}}.
\eeq
Here $d_{BG}$ is the distance of the back gate from where the 
two- dimensional electron gas, (2 DEG), is. In GaAs, $\epsilon $ 
=13.1 and $d_{BG}$= 428 $\pm $ 5 $\mu$ m. From (8), we find the 
charge and substitute the value of $C$ from (9) to obtain,
\beg
q= \Delta V_{BG} C S_m= \Delta V_{BG} S_m{\epsilon \epsilon_o\over
 d_{BG}}.
\eeq
Now write (5) as $S_m \Delta B=\phi_o/p_\nu$ with an extra number 
$p_\nu$ and substitute this value of $S_m$ in (10) to obtain,
\beg
q=\Delta V_{BG}S_m{\epsilon \epsilon_o\over d_{BG}}=\Delta
 V_{BG}{\epsilon\epsilon_o\over d_{BG}}{\phi_o\over p_\nu\Delta B}
\eeq
so that the charge becomes,
\beg
q={\epsilon \epsilon_o \phi_o\Delta V_{BG}\over p_\nu d_{BG}
 \Delta B}.
\eeq
When $p_\nu$ was introduced, it must be the inverse of an integer 
so that as long as $p_\nu$ =1, there is no trouble at all but 
$p_\nu$  can not be 2 because 2 is not the inverse of a number 
but 1/2 can be a component in (n+1/2)=$1/p_\nu$ so that when 
$n$=0, $p_\nu$=2 is not much objectionable. So there is no objection 
to $p_\nu$=1 or 2. However, it is clear that eq.(12) measures the 
product $q\Delta B$ and not the charge, $q$, and this is a serious
 matter.

     We conclude that the experiment of Goldman and Su does not
 measure the charge of a quasi-particle but it measures the product
 of charge and a magnetic field. Let us write the field as the 
inverse area. Then the experiment measures the charge per unit area
 and not the charge,
\beg
{e^*\over area^*}{area\over e}= 0.331.
\eeq
This is the quantity measured and it is claimed that,
\beg
{e^*\over e}= 0.331.
\eeq
Therefore Goldman and Su's experiment does not measure the 
fractional charge. The area in the above is $\phi_o/B$. Hence 
the quantity measured is the product of charge and the magnetic 
field, $B$.

     {\bf (b)}: de-Picciotto et al [4] report that a 
two-dimensional electron gas, subjected to a strong perpendicular
 magnetic field, B, consists of highly degenerate Landau levels 
with a degeneracy per unit area,
\beg
 p= B/\phi_o
\eeq
with 
\beg
\phi_o=h/e
\eeq
the flux quantum with $h$ as the Planck's constant. Actually
 in both these expressions, the dimensions are incorrect. The 
correct formula for $p$ is,
\beg
p=B.A/\phi_o
\eeq
and the correct value of $\phi_o$ is,
\beg
\phi_o={hc\over e}.
\eeq
This is not a serious error because it can be corrected 
when necessary.
The area $A$ is $p\phi_o/B$ which defines the magnetic length,
 $\sqrt A=(p\phi_o/B)^{1/2}$. The areal density is $n_s$, which 
is the number of electrons per unit area,
\beg
n_s=n/A.
\eeq
The integer number $\nu$ is the filling factor,
\beg
\nu =n_s/p.
\eeq
This is the number per unit area, so it can not be correct. 
The correct value is obtained only when $p$ is devided by the
 area $A$ so that,
\beg
\nu = {n_s.A\over p}
\eeq
is dimensionless. The Hall resistivity is defined as,
\beg
\rho_{xy} ={h\over \nu e^2}
\eeq
and hence the conductivity,
\beg
\sigma_{xy} ={\nu e^2\over h}
\eeq
is correct with $\nu$ as a number. It has been argued that the 
resistivity observed in the fractional quantum Hall effect, can be 
explained in terms of quasiparticles of a fractional charge,
\beg
Q={e\over q}\,\,\, (q=number).
\eeq
We have no objection to this charge equation where $q$ is a number
 but it will be interesting to see that there are other factors in 
the problem so that even if the fractional charge is observed, it 
does not necessarily mean that there is a fractional charge. Such a 
fraction can arise from the angular  momentum so that the effective 
charge appears to be fractional. This means that the observed charge
 is fractional but the fraction can arise without fractionalization 
of charge. De Picciotto et al suggest that `` the FQH effect cannot 
be explained and is believed to result from interactions between the 
electrons brought about by the strong magnetic field". In this case 
the ``electrons brought about by the strong magnetic field" may be 
correct but this is not born out from
Laughlin's paper. De Picciotto et al have argued against the paper 
of Goldman and Su. We have seen above that their experiment measures 
the product of the magnetic field and the charge but not charge alone.
 De Picciotto et al report that `` Quantum shot noise, probes the 
temporal behaviour of the current and thus offers a direct way to 
measure the charge". We wish to examine whether it is correct to say 
that `` quantum shot noise" measures the charge. Actually, the noise
 measures the resistivity not the charge. The fractional filling 
factor is the inverse of the fraction which measures the charge as 
in $Q$ (eq.11), so that,
\beg
\nu={1\over q}.
\eeq
The thermal noise is,
\beg
S_i= 4k_BTG
\eeq
where $G$ is the conductance, so that a plot of $S_i$ versus $G$ 
determines the temperature. The quantum shot noise, $S_i$, 
generated by the backscattering of the current, $I_B$, is 
proportional to the charge of the quasiparticle so that,
\beg
S_i=2QI_B
\eeq
The magnetic field is swept from $0$ to 14 Tesla, which means
 that product of field and charge is measured but not the charge. 
The transmission, $t$, is given by the ratio of conductance $G$ 
and the quantum conductance $g_o$=$e^2/h(area^\prime)$ as,
\beg
t= ({G\over area}).({area^\prime\over g_o}).
\eeq
The charge $Q=e/3$, when measured from the transmission, $t$, 
without regard to areas involved, will not be the true charge 
of the quasiparticles. Besides, there is nonlinearity in the 
$I-V$ characteristics but only a small voltage range is shown.

{\bf (c)}:Let us look at the measurements performed by Saminadayar 
et al[5]. It has been reported that 2-dimensional electrons in a 
high magnetic field give rise to Landau levels with one state
 per flux,
\beg
\phi_o= { hc\over e}.
\eeq
For integer filling factor, $\nu$=$n_s/n_\phi$,
\beg
n_\phi ={eB\over hc}.
\eeq
Here it is obvious that if $n_\phi$ is dimensionless an area is 
missing.
Therefore, $n_\phi$ should be described by an area,
\beg
{n_\phi\over A}= {eB\over hc}.
\eeq
This is the corrected form of the previous expression. So the
 flux density is $n_\phi/A$, which is the number of flux quanta 
per unit area. Similarly, $n_s$ is the electron density. If the 
distance between two electrons is equal to the magnetic length, 
the Coulomb energy $\Delta$=$e^2/\epsilon l_c$ is called a gap with,
\beg
{\it l}^2_c={\hbar c\over eB}.
\eeq
According to Ohm's law, the current $I_B$ is related to the 
voltage, $V_{ds}$ by the relation,
\beg
V=I.{h\over \nu e^2}.
\eeq
If $\nu$ is the effective charge, the cyclotron frequency is 
given by,
\beg
\omega = {\nu eB\over mc}
\eeq
in which we substitute $B=(hc/eA)$ to find,
\beg
\omega={\nu h\over  mA}.
\eeq
The quantity being measured on the right hand side is $\nu/A$ but
 not $\nu$ alone. The previous formula shows that a product 
$\nu eB$ occurs
so that whether $\nu$ is to multiply $e$ or $B$ can not be 
ascertained. It is stated that ``we have brought evidence of $e/3$ 
Laughlin's[6] quasiparticles carrying current through the 1/3 FQH 
state. However, a
close examination [7 ] of Laughlin's original paper shows that the
 quantity that appears in the algebra is $e/(ma_o^2)$ where $m=3$ 
and $a_o=1$ gives $e/3$ used by Saminadayar et al. However, $a_o$ 
is not a dimensionless number and $a_o=\hbar c/eB$ which depends
 on the magnetic field so that what is thought to be $e/3$ is 
actually $eB/3$ so that whether $e/3$ or $B/3$ has been measured, 
can not be determined.

{\bf (d)}: Laughlin[6] has found the charge density as explained 
by us [7],
\beg
\rho = {e\over m 2\pi a_o^2}.
\eeq
If $e=x$ and $a_o=y$ then the above equation is,
\beg
2\pi m \rho y^2=x.
\eeq
Laughlin substituted $y=1$ and then determined the value of $x=e/3$.
 However, y=1 is not a solution and this arbitrarily fixing $y=1$ 
called
``incompressibility" is not justified. Since there is only one 
equation and these two variables, Laughlin could not solve this 
equation correctly. Obtaining the value of $x/y^2$ exactly is not 
the solution of the problem. So the exactness is of no help and
 hence what is an exact solution is not necessarily satisfactory.

     Similarly, the experiments of Comforti et al [8,9] require
 that three quasiparticles must group so that the total charge is 
$e$ and 1/3 is prepared in a field only. So the quantity measured
 is the field, not the charge.

     Anderson[10] has suggested a state with extra unit of angular 
momentum. An effort is made to create quasiparticles of charge 2/m 
and 3/m, etc. from a modification of Laughlin's state. From a
 different approach we find that the charge of the quasiparticles is,
\beg
e_{eff}/e={{\it l} +(1/2)\pm s \over (2{\it l}+1)}.
\eeq
Thus the quasiparticle charge depends on spin, s, and there is an
electrically neutral quasiparticle which has a spin of 1/2.

     The proper theory of the quantum Hall effect is given in Ref.11
which agrees with Stormer's data[12].

\noindent{\bf3.~~ Conclusions}.

     It will be best if Laughlin's theory is interpreted not as
 determining the charge but determining $e/a_o^2$. There is not 
much
harm in measuring the fractional charge but the correct result
 helps in determining the origin of the factor of $1/3$. All of 
the experiments prepare the 1/3 charge in the presence of a 
magnetic field and this field is always needed.

\noindent{\bf4.~~References}
\begin{enumerate}
\item V. J. Goldman and B. Su, Science, {\bf 267}, 1010(1995).
\item V. J. Goldman, Physica E {\bf 1}, 15 (1997).
\item V. J. Goldman, cond-mat/9708041
\item R. de-Picciotto, M. Reznikov, M. Heiblum, V. Umansky,
 G. Bunin and D. Mahalu, Nature, {\bf 389}, 162(1997);
Physica B{\bf 249}, 395 (1998).
\item L. Saminadayar, D. C. Glattli, Y. Jin and B. Etienne,
 Phys. Rev. Lett. {\bf 79}, 2526 (1997).
\item R. B. Laughlin, Phys. Rev. Lett. {\bf 50}, 1395 (1983).
\item K. N. Shrivastava, cond-mat/0212552.
\item E. Comforti, Y. C. Chung, M. Heiblum, V. Umansky and 
D. Mahalu, Nature, {\bf 416}, 515 (2002).
\item E. Comforti, Y. C. Chung, M. Heiblum and V. Umansky,
 cond-mat/0112367.
\item P. W. Anderson, Phys. Rev. B {\bf 28}, 2264 (1983).
\item K.N. Shrivastava, Introduction to quantum Hall effect,\\ 
      Nova Science Pub. Inc., N. Y. (2002).
\item H. L. Stormer, Rev. Mod. Phys. {\bf 71}, 875 (1999).
\end{enumerate}

{\it Shrivastava has worked with Roy Anderson, Ken Stevens,
 Vincent Jaccarino and others. His work has been useful to 
M. C. R. Symons, F.R.S.,  and to K. Alex M\"uller. One of his 
papers was communicated to the Proc. Roy. Soc. by B. Bleaney,
 F.R.S. He has worked at Harvard University, University of 
Houston, University of California at Santa Barbara, etc.He has 
published 170 papers including two books.}

{\bf Note:} Ref.11 is available from:
 Nova Science Publishers, Inc.,
400 Oser Avenue, Suite 1600,
 Hauppauge, N. Y. 11788-3619,
Tel.(631)-231-7269, Fax: (631)-231-8175,
 ISBN 1-59033-419-1 US$\$69$.\\
E-mail: novascience@Earthlink.net

\end{document}